\documentclass[preprint]{aastex}

\begin{document}

\title {$\beta$ Cephei Pulsations in the High-mass Eclipsing System CW Cephei }
\author{Jae Woo Lee$^{1}$ and Kyeongsoo Hong$^{2}$}
\affil{$^1$Korea Astronomy and Space Science Institute, Daejeon 34055, Republic of Korea}
\affil{$^2$Institute for Astrophysics, Chungbuk National University, Cheongju 28644, Republic of Korea}
\email{jwlee@kasi.re.kr}

\begin{abstract}
CW Cep is an early B-type eclipsing binary with mass measurement precisions better than 1 \%. We report the discovery of 
pulsation signatures in the TESS time-series data of the system observed during Sectors 17 and 18. Our binary modeling indicates 
that the target star is a partially-eclipsing detached system with masses of 12.95 $M_\odot$ and 11.88 $M_\odot$ and radii of 
5.52 $R_\odot$ and 5.09 $R_\odot$ in an eccentric orbit of $e$ = 0.0305. The distance to the eclipsing system, 928 $\pm$ 36 pc, 
is much more precise than the Gaia distance of 962 $\pm$ 453 pc. Applying multifrequency analyses to the residual light curve in 
the outside-eclipse part, we detected 13 significant signals in two frequency regions. Six frequencies below 1 day$^{-1}$ appeared 
to be mostly orbital harmonic and combination terms, or sidelobes due to insufficient removal of the binary effects. In contrast, 
seven frequencies clustered around 2.73 day$^{-1}$ and 5.34 day$^{-1}$ could be considered $\beta$ Cep-type pulsations. 
Our results represent the second discovery of $\beta$ Cep pulsations present in double-lined eclipsing binaries with precise masses 
and, hence, CW Cep serves as an important test-bed for the asteroseismic modeling of high-mass stars.  
\end{abstract}


\section{INTRODUCTION}

The study of stellar structure and evolution is based on precisely measured absolute parameters. Double-lined eclipsing binaries 
(EBs) offer us the reliable masses and radii of each star from time-series spectroscopy and photometry (Hilditch 2001), 
whose precisions are below 1 \% for best-studied EBs (Torres et al. 2010; Lee et al. 2020b). Such measurements are widely used to 
improve theoretical stellar models and to determine the distances to EB systems. Although high-mass stars (8 $\ga$ $M_\odot$) are 
uncommon compared to lower-mass ones, they are powerful cosmic engines in the Universe (Langer 2012). However, there are often 
significant differences between the masses derived from the spectroscopic analyses and the stellar evolution models: 
the former masses are systematically smaller than the latter masses (Herrero et al. 1992; Tkachenko et al. 2020). This issue is 
called the mass discrepancy problem, and indicates how important it is to determine the dynamical masses of high-mass EBs. 

Stellar pulsations can occur at almost every stage of evolution on the Hertzsprung-Russell diagram. The pulsational characteristics 
of oscillating stars provide valuable information on stellar interior structure. For example, $\beta$ Cep-type variables are 
some of the most massive and hot pulsators. They are early B-type stars (B0$-$B2.5) and exhibit light, radial velocity (RV), and/or 
line profile variations typically over periods of several hours (Stankov \& Handler 2005; Labadie-Bartz et al. 2020). These changes 
are due to low-order pressure and gravity pulsations excited by the $\kappa$ mechanism (Moskalik \& Dziembowski 1992; 
Dziembowski \& Pamiatnykh 1993). The photometric pulsation amplitudes are typically smaller than 0.1 mag and very similar, from blue 
to infrared bandpasses (Stankov \& Handler 2005). $\beta$ Cep pulsating stars in EBs are prime objects for studying the internal 
structure and evolution of high-mass stars between 8 $M_\odot$ and 17 $M_\odot$ using binary properties and asteroseismic techniques. 
At present, the high-mass star V453 Cyg is the only double-lined EB known to have a $\beta$ Cep pulsating component 
(Pigulski \& Pojma\'nski 2008; Ratajczak et al. 2017; Southworth et al. 2020). 

In this study, we found $\beta$ Cep-type pulsations with amplitudes lower than 0.5 mmag in the B-type eclipsing star CW Cep using 
highly precise TESS data. The program target (TIC 434893323; HD 218066; TYC 4282-419-1; $T_{\rm p}$ = $+$7.234; $V\rm_T$ = $+$7.728, 
$(B-V)\rm_T$ = $+$0.352) is known to be a double-lined EB, with both an apsidal motion and a light time effect due to a circumbinary 
object (Han et al. 2002). Previous studies have shown that the primary star being eclipsed at Min I (orbital phase 0.0) is 
more massive and larger than the secondary component, and that a third light ($l_3$) is needed to model the eclipsing light curves. 
The $l_3$ contributions to the EB system have been found to be over a broad range of 0.6$-$9.7 \%, which is greater at longer 
wavelengths (Clausen \& Gimenez 1991; Han et al. 2002; Johnston et al. 2019). A detailed literature review of the CW Cep works 
appears in Johnston et al. (2019). They obtained orbital parameters and effective temperatures from their high-resolution spectroscopy 
and spectral disentangling, and derived the fundamental stellar parameters ($M_1$ = 13.00 M$_\odot$, $M_2$ = 11.94 M$_\odot$, 
$R_1$ = 5.45 R$_\odot$, and $R_2$ = 5.09 R$_\odot$) with a per-cent level of precision by applying them to the $uvby$ light curves 
of Clausen \& Gimenez (1991). Here, the subscripts 1 and 2 refer to the primary and secondary stars, respectively.

\section{OBSERVATIONS AND ECLIPSE TIMINGS}

For our study, we used the 2-min sampling data of CW Cep available from the TESS mission (Ricker et al. 2015). The observations were 
taken during Sectors 17 and 18 (BJD 2,458,764.67 $-$ 2,458,815.04) and downloaded from the Mikulski Archive for Space Telescopes (MAST). 
The raw SAP light curve was detrended and normalized by individually fitting a second-order polynomial to four segments separated by 
data gaps. Then, the normalized fluxes were converted to magnitude scales using the TESS magnitude of $T_{\rm p}$ = $+$7.234 
(Stassun et al. 2019). This procedure was the same as that applied by Lee et al. (2017, 2019). The resulting time-series data are 
displayed as magnitude versus BJD in Figure 1. 

We measured 31 eclipse timings from the TESS photometric observations during both minima (Min I and Min II). The minimum epochs and 
their uncertainties in Table 1 were computed using the method of Kwee \& van Woerden (1956). To obtain the phase-folded light curve 
of CW Cep, we applied a least-squares fit to all 16 primary minima and resulted in the following linear ephemeris: 
\begin{equation}
 \mbox{Min I} = \mbox{BJD}~ (2458781.2997514\pm0.0000072) + (2.7291321\pm0.0000011)E.
\end{equation}
The binary period corresponds to the orbital frequency of $f_{\rm orb}$ = 0.3664169 $\pm$ 0.0000002 day$^{-1}$. The TESS light curve 
using this equation is presented in Figure 2.

\section{BINARY MODELING}

The ultra-precise TESS observations of CW Cep were modeled by applying the detached mode 2 of the Wilson-Devinney (W-D) program 
(Wilson \& Devinney 1971; van Hamme \& Wilson 2007), in a manner almost identical to that for the TESS pulsating eclipsing star AI Hya 
(Lee et al. 2020a). The binary mass ratio is of paramount importance in the W-D modeling because the Roche geometry is quite sensitive 
to this parameter. In this modeling, we set the mass ratio and the surface temperature of the more massive, hotter primary component 
to be $q$ = 0.917 $\pm$ 0.002 and $T_{\rm eff,1}$ = 28,300 $\pm$ 460 K, respectively, updated from the high-quality echelle spectra 
of Johnston et al. (2019). The bolometric albedos of $A_{1,2}$ = 1.0 and the gravity-darkening coefficients of $g_{1,2}$ = 1.0 were 
used, and the logarithmic limb-darkening coefficients ($X_{\rm bol}$, $Y_{\rm bol}$, $x_{T_{\rm P}}$, and $y_{T_{\rm P}}$) were found 
by interpolating the values of van Hamme (1993). Most of the other parameters were initialized from Han et al. (2002) and Johnston et al. (2019). 

Our synthesis was carried out in two stage. In the first stage, we fitted the following light curve parameters: the ephemeris epoch 
($T_0$) and period ($P$), the argument of periastron ($\omega$), the orbital eccentricity ($e$) and inclination ($i$), the secondary's 
temperature ($T_2$), the dimensionless surface potentials ($\Omega_1$, $\Omega_2$), and the monochromatic luminosity ($L_1$). 
The TESS light curve was repeatedly analyzed until the corrections to the fitted parameters were smaller than their uncertainties. 
The binary solution appears as a green curve in Figure 2. The light residuals from it are given in the middle panel to show 
the unmodeled light variations. The model curve does not describe the observed TESS data satisfactorily, including both of 
the eclipse minima. 

In the second stage, we calculated a series of models for each assumed $l_3$ between 0.0 and 0.2. The $l_3$ search indicates 
a global minimum of the weighted sum of the squared residuals, $\sum{W(O-C)^2}$, at $l_3$ = 0.12. We also tested the $A_{1,2}$ and 
$g_{1,2}$ parameters, but their fits do not have much impact on the light curve residuals. Thus, the three parameters of $l_3$, 
$x_{T_{\rm P_{1}}}$, and $x_{T_{\rm P_{2}}}$ were added as adjustable variables. The consequential parameters are presented in Table 2. 
To obtain the parameter errors, we ran various models with different approaches for mass ratio, temperature, eccentricity, third light, 
limb darkening, albedo, and gravity darkening (Southworth et al. 2020). The parameter errors in Table 2 are estimated from 
the differences between our final solution and the models considered. The synthetic curve from the Table 2 parameters appears as 
a red curve in Figure 2, and the corresponding residuals are plotted in its bottom panel. We can see that the binary model including 
$l_3$ and $x_{T_{\rm P_{1,2}}}$ significantly improves the light curve fitting.

\section{PULSATIONAL CHARACTERISTICS}

It is possible that both components of CW Cep may be $\beta$ Cep variables, based on their fundamental parameters (Pamyatnykh 1999; 
Stankov \& Handler 2005). At first glance, the light residuals from our binary modeling seem to include information on the possibility 
of oscillations with an amplitude of about 5 mmag. For better multifrequency analyses, we split the TESS time-series data of CW Cep 
into an interval of an eclipsing period ($P$), and individually analyzed the 17 light curves by fitting only $T_0$ among the binary 
parameters in Table 2. Figure 3 presents the combined residuals from this process as magnitude versus BJD. 

During eclipses, both components of CW Cep partially block each other's lights, which can make the multifrequency analyses 
difficult. To avoid such a problem and search for possible periodic variations up to the Nyquist limit of 360 day$^{-1}$, 
we introduced the outside-eclipse light residuals (orbital phases 0.072$-$0.414 and 0.554$-$0.928) into the \textsc{PERIOD04} program 
(Lenz \& Breger 2005). Using the consecutive prewhitening method (Lee et al. 2014), we extracted a total of 13 significant frequencies 
presented in Table 3, where the parameters' errors were computed following Kallinger et al. (2008). No prominent signals were detected 
in the frequency region higher than 6 day$^{-1}$. The Fourier amplitude spectra for CW Cep are displayed in Figure 4. The dotted line 
in the third panel denotes the four times the noise spectrum was computed over an interval of 0.1 day$^{-1}$ (Breger et al. 1993). 
The synthetic curve from the detected frequencies is presented as a red line in the lower panel of Figure 3. 

As shown in Table 3 and Figure 4, the dominant frequency signals of CW Cep lie in two ranges of 2$-$6 day$^{-1}$ and $<$ 1 day$^{-1}$. 
The observation time span of CW Cep is $\Delta T$ = 49.8 days. Within the frequency resolution of 1.5/$\Delta T$ = 0.030 day$^{-1}$, 
we looked for orbital harmonics ($nf_{\rm orb}$) and combination terms ($nf_i \pm mf_j$). Here, $n$ and $m$ are integer numbers, and 
$f_i$ and $f_j$ are parent frequencies. The search result is given in the last column of Table 3. Frequencies $f_1$ and $f_5$ are 
the orbital frequency and its multiples, respectively, and $f_{9}$, $f_{10}$, and $f_{11}$ likely come from combination frequencies. 
On the other hand, the high-precision TESS data of CW Cep are not fully fitted by our binary modeling, and the detected frequencies 
could be affected by the unmodeled light residuals. Thus, to check above frequency analysis, we first formed 1,000 normal points in 
a bin width of phase 0.001. Then, the outside-eclipse residuals from them were analyzed in the same way as before, and the resultant 
spectrum was plotted. It is presented in the lowest panel of Figure 4. The frequencies other than $<$ 1 days$^{-1}$ matched well with 
those from the binary modeling. The frequencies between 2 day$^{-1}$ and 6 day$^{-1}$ are typical of $\beta$ Cep pulsations 
(Stankov \& Handler 2005; Southworth et al. 2020).

\section{DISCUSSION AND CONCLUSIONS}

In this paper, we presented the TESS time-series data for CW Cep, observed in two sectors 17 and 18, to find pulsation features 
in the early B-type eclipsing star. Our binary model fit shows that the program target is a partially-eclipsing detached system 
in an elliptical orbit of $e$ = 0.0305, as reported in previous studies. By conducting the $l_3$ search in our modeling process, 
we recognized that the third light contributes about 12\% to the entire luminosity of the CW Cep system. The $l_3$ source may be 
mainly a physically-bound invisible circumbinary companion, and it could be partly affected by the existence of a nearby star 
TIC 434893308 (RA$_{2000}$ = 23$^{\rm h}$03$^{\rm m}$59$\fs715$; DEC$_{2000}$ = +63$^{\circ}$24${\rm '}$02$\farcs$22; $T_{\rm p}$ = $+$11.67) 
with a separation of about 21 arcsec. The filling factors ($F_{1,2}$ = $\Omega_{\rm in}$/$\Omega _{1,2}$) of the component stars are 
$F_1$ = 68 \% and $F_2$ = 67 \%, respectively, and the ratio of their fractional radii is $r_2$/$r_1$ = 0.9219. 

We derived the fundamental stellar parameters of CW Cep by combining our binary solution and the RV amplitudes ($K_1$ = 211.1 $\pm$ 0.4 km s$^{-1}$ 
and $K_2$ = 230.2 $\pm$ 0.4 km s$^{-1}$) of Johnston et al. (2019). The result is presented in the bottom part of Table 2, where 
the bolometric corrections (BCs) are taken from the $\log T_{\rm eff}$$-$BC relation of Torres (2010). For the distance determination, 
first of all we obtained the apparent magnitude of $V$ = +7.70 $\pm$ 0.01 and the color index of ($B-V$) = 0.30 $\pm$ 0.02 from 
$B\rm_T$ and $V\rm_T$ in the Tycho-2 Catalogue (H\o g et al. 2000). Then, $E$($B-V$) = $+$0.58 was estimated from the color difference 
between the observed ($B-V$) and the intrinsic ($B-V$)$\rm_0$ = $-$0.28 for the primary star's temperature (Flower 1996). Finally, 
the distance to CW Cep was computed to be 928 $\pm$ 36 pc, which is much more precise than that (962 $\pm$ 453 pc) by the Gaia parallax 
of 1.04$\pm$0.49 mas (Gaia Collaboration et al. 2018).

Multifrequency analyses were applied to all of the residuals of CW Cep in the outside-eclipse orbital phases. As a consequence, 
we found 13 significant signals, and most of the frequencies below 1 day$^{-1}$ might be identified as possible sidelobes, due to 
either imperfect binary modeling or trends present in the TESS data. In contrast, the seven frequencies at intervals of 
2$-$6 day$^{-1}$ are strongly reminiscent of $\beta$ Cep pulsations. Among them, at least two ($f_2$ and $f_3$) with high signal 
to noise amplitude ratio (S/N) are independent pulsation frequencies, and correspond to pulsation periods of 8.803 h and 4.494 h, 
respectively. However, at present, we do not know which of the component stars exhibit $\beta$ Cep pulsations. 
High-resolution time-series spectroscopy will help identify a pulsating component in the CW Cep system by detecting the RV and/or 
line-profile variations caused by pulsation modes. The results from our binary model and frequency analysis reveal that $\beta$ Cep-type 
pulsations are present in the high-mass eclipsing star CW Cep. Including our target star, two high-mass double-lined EBs are known 
to show $\beta$ Cep pulsations (Southworth et al. 2020). Thus, the pulsating EB CW Cep would provide a crucial benchmark for 
the asteroseismic study of this important stellar group.

\acknowledgments{ }
This paper includes data collected by the TESS mission, which were obtained from MAST. Funding for the TESS mission is provided by 
the NASA Explorer Program. The authors wish to thank the TESS team for its support of this work. We also appreciate the careful reading 
and valuable comments of the anonymous referee. This research has made use of the Simbad database maintained at CDS, Strasbourg, France, 
and was supported by the KASI grant 2020-1-830-08. K.H. was supported by the grants 2017R1A4A1015178 and 2019R1I1A1A01056776 from 
the National Research Foundation (NRF) of Korea.

\clearpage
\begin{figure}
\includegraphics{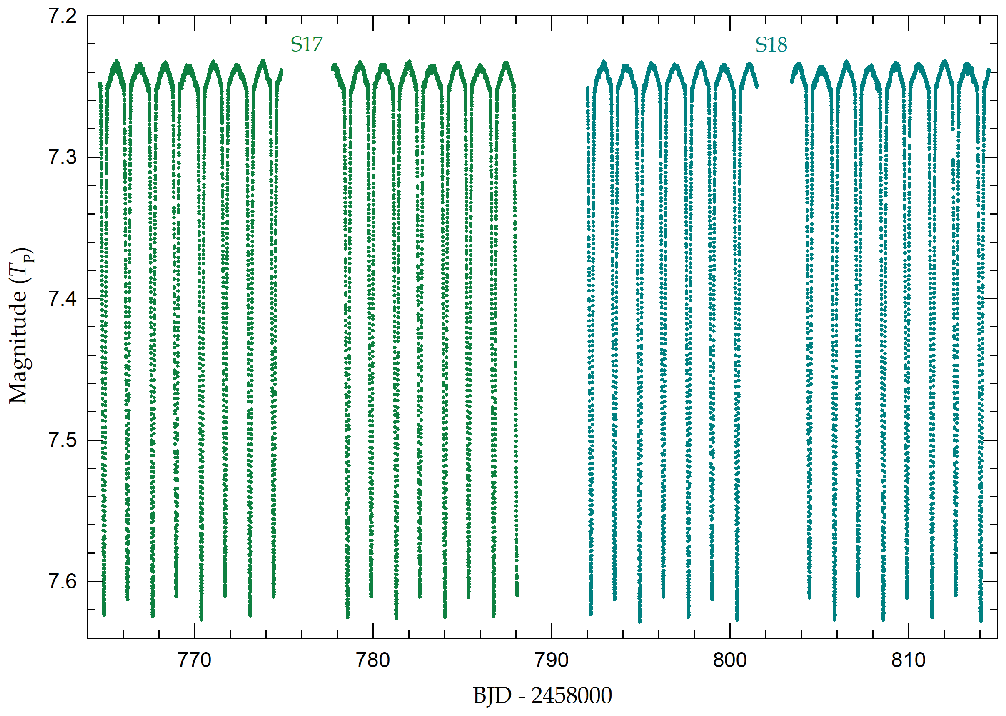}
\caption{TESS time-series data of CW Cep observed during Sectors 17 (green) and 18 (cyan). }
\label{Fig1}
\end{figure}

\begin{figure}
\includegraphics[]{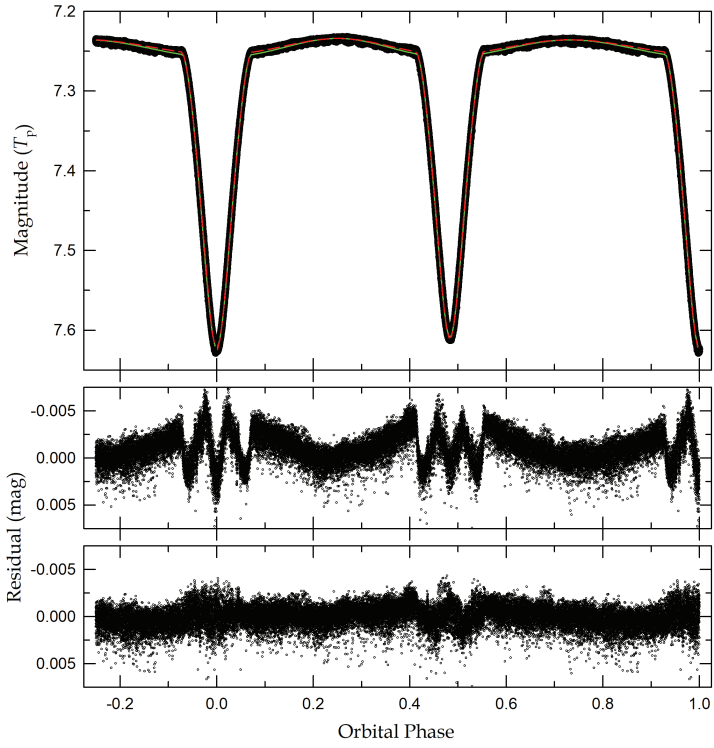}
\caption{Phase-folded light curve of CW Cep with the fitted models. In the top panel, the green and red lines represent 
the synthetic curves obtained from our binary models: without and with the $l_3$ and $x_{T_{\rm P_{1,2}}}$ parameters. 
The corresponding residuals from the two models are plotted in the middle and bottom panels, respectively. }
\label{Fig2}
\end{figure}

\begin{figure}
\includegraphics[]{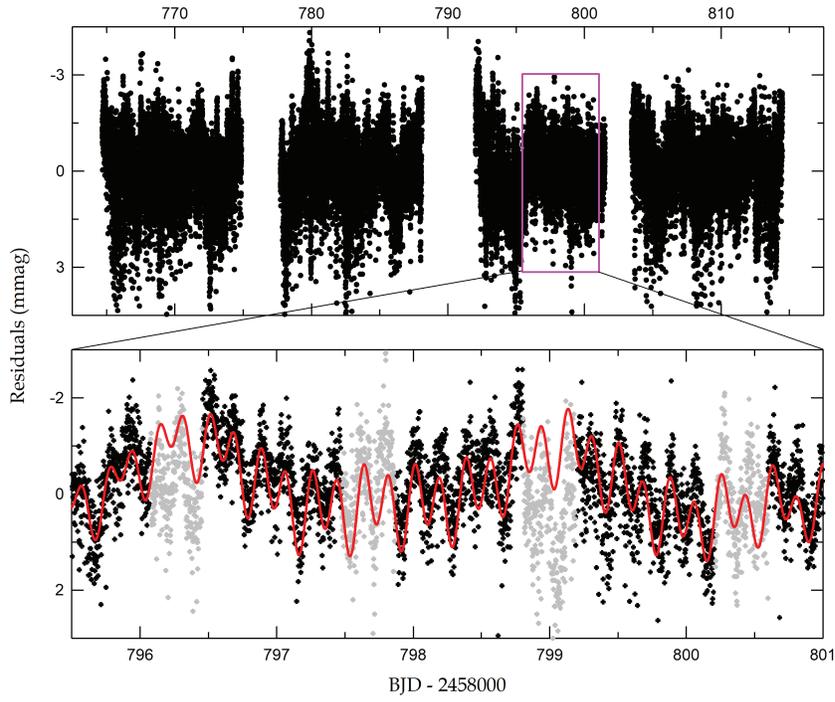}
\caption{Light curve residuals distributed in BJD. The lower panel presents a short section of the residuals marked using the inset box 
in the upper panel. The gray circles indicate the times of the primary and secondary eclipses, and the synthetic curve is computed from 
the 13-frequency fit to the out-of-eclipse data. }
\label{Fig3}
\end{figure}

\begin{figure}
\includegraphics[]{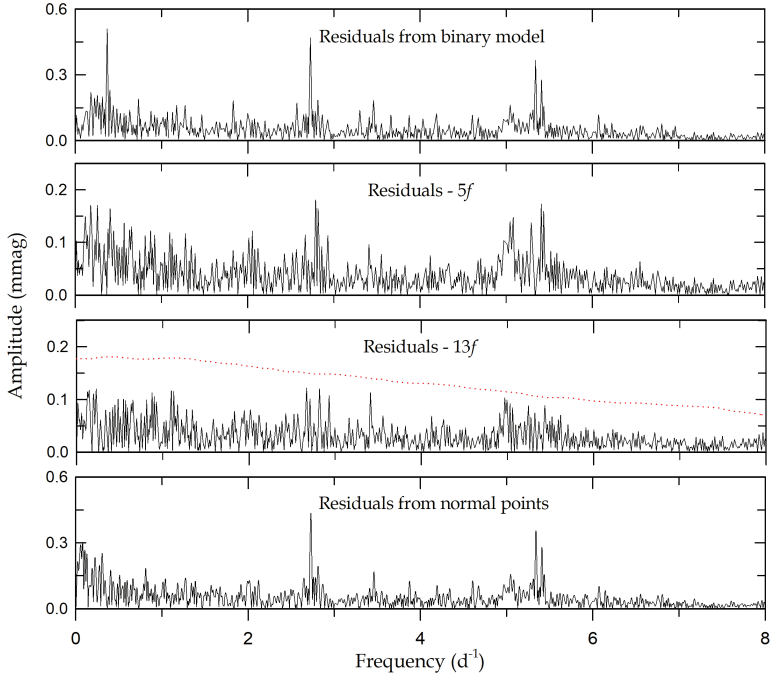}
\caption{Amplitude spectra before (top panel) and after pre-whitening the first five frequencies (second) and all 13 frequencies (third) 
detected by analyzing the outside-eclipse residuals from the best-fit binary model. The dotted line in the third panel corresponds to 
four times the noise spectrum, which was calculated for each frequency in an equidistant step of 0.1 day$^{-1}$. The periodogram for 
the light residuals from 1000 normal points is shown in the bottom panel for comparison. }
\label{Fig4}
\end{figure}

\clearpage                                                                                                           
\begin{deluxetable}{lccclcc}
\tablewidth{0pt}
\tablecaption{TESS Eclipse Timings for CW Cep }
\tablehead{
\colhead{BJD$_{\rm TDB}$} & \colhead{Error} & \colhead{Min} & \colhead{BJD$_{\rm TDB}$} & \colhead{Error} & \colhead{Min}
}
\startdata
2,458,764.924891  & $\pm$0.000020   & I             & 2,458,792.216451  & $\pm$0.000040   & I       \\
2,458,766.244292  & $\pm$0.000051   & II            & 2,458,793.535757  & $\pm$0.000048   & II      \\
2,458,767.654091  & $\pm$0.000040   & I             & 2,458,794.945412  & $\pm$0.000033   & I       \\
2,458,768.973275  & $\pm$0.000044   & II            & 2,458,796.265006  & $\pm$0.000027   & II      \\
2,458,770.383285  & $\pm$0.000021   & I             & 2,458,797.674509  & $\pm$0.000033   & I       \\
2,458,771.702702  & $\pm$0.000021   & II            & 2,458,798.994269  & $\pm$0.000035   & II      \\
2,458,773.112280  & $\pm$0.000032   & I             & 2,458,800.403801  & $\pm$0.000035   & I       \\
2,458,774.431510  & $\pm$0.000033   & II            & 2,458,804.452804  & $\pm$0.000029   & II      \\
2,458,778.570286  & $\pm$0.000036   & I             & 2,458,805.862075  & $\pm$0.000022   & I       \\
2,458,779.890270  & $\pm$0.000086   & II            & 2,458,807.181401  & $\pm$0.000027   & II      \\
2,458,781.299927  & $\pm$0.000023   & I             & 2,458,808.590669  & $\pm$0.000039   & I       \\
2,458,782.618884  & $\pm$0.000038   & II            & 2,458,809.910896  & $\pm$0.000029   & II      \\
2,458,784.028799  & $\pm$0.000023   & I             & 2,458,811.320216  & $\pm$0.000020   & I       \\
2,458,785.348485  & $\pm$0.000018   & II            & 2,458,812.639674  & $\pm$0.000032   & II      \\
2,458,786.758070  & $\pm$0.000019   & I             & 2,458,814.049127  & $\pm$0.000034   & I       \\
2,458,788.077753  & $\pm$0.000303   & II            &                   &                 &         \\
\enddata
\end{deluxetable}

\begin{deluxetable}{lcc}
\tabletypesize{\small}
\tablewidth{0pt} 
\tablecaption{Binary Parameters of CW Cep }
\tablehead{
\colhead{Parameter}               & \colhead{Primary}  & \colhead{Secondary}                                                  
}                                                                                                                                     
\startdata                                                                                                                            
$T_0$ (BJD$_{\rm TDB}$)           & \multicolumn{2}{c}{2,458,781.276886$\pm$0.000026}             \\
$P$ (day)                         & \multicolumn{2}{c}{2.7291428$\pm$0.0000027}                   \\
$q$                               & \multicolumn{2}{c}{0.917$\pm$0.002}                           \\
$e$                               & \multicolumn{2}{c}{0.03051$\pm$0.00086}                       \\
$\omega$ (deg)                    & \multicolumn{2}{c}{212.7$\pm$3.2}                             \\
$i$ (deg)                         & \multicolumn{2}{c}{82.947$\pm$0.058}                          \\
$T$ (K)                           & 28,300$\pm$460               & 27,431$\pm$434                 \\
$\Omega$                          & 5.320$\pm$0.025              & 5.418$\pm$0.022                \\
$\Omega_{\rm in}$$\rm ^a$         & \multicolumn{2}{c}{3.614}                                     \\
$X_{\rm bol}$, $Y_{\rm bol}$      & 0.722, 0.196                 & 0.731, 0.185                   \\
$x_{T_{\rm P}}$, $y_{T_{\rm P}}$  & 0.219$\pm$0.038, 0.198       & 0.213$\pm$0.036, 0.194         \\
$l$/($l_{1}$+$l_{2}$+$l_{3}$)     & 0.4875$\pm$0.0046            & 0.3910                         \\
$l_{3}$$\rm ^b$                   & \multicolumn{2}{c}{0.1215$\pm$0.0043}                         \\
$r$ (pole)                        & 0.2273$\pm$0.0014            & 0.2099$\pm$0.0012              \\
$r$ (point)                       & 0.2354$\pm$0.0016            & 0.2166$\pm$0.0014              \\
$r$ (side)                        & 0.2299$\pm$0.0014            & 0.2120$\pm$0.0013              \\
$r$ (back)                        & 0.2338$\pm$0.0015            & 0.2153$\pm$0.0013              \\
$r$ (volume)$\rm ^c$              & 0.2305$\pm$0.0015            & 0.2125$\pm$0.0013              \\ [1.0mm] 
\multicolumn{3}{l}{Absolute parameters:}                                                          \\
$M$ ($M_\odot$)                   & 12.951$\pm$0.052             & 11.877$\pm$0.049               \\
$R$ ($R_\odot$)                   & 5.520$\pm$0.037              & 5.089$\pm$0.032                \\
$\log$ $g$ (cgs)                  & 4.066$\pm$0.006              & 4.099$\pm$0.006                \\
$\rho$ ($\rho_\odot$)             & 0.077$\pm$0.002              & 0.090$\pm$0.002                \\
$\log$ $L$ ($L_\odot$)            & 4.243$\pm$0.029              & 4.119$\pm$0.028                \\
$M_{\rm bol}$ (mag)               & -5.879$\pm$0.072             & -5.566$\pm$0.070               \\
BC (mag)                          & -2.727$\pm$0.044             & -2.643$\pm$0.042               \\
$M_{\rm V}$ (mag)                 & -3.152$\pm$0.085             & -2.923$\pm$0.082               \\
Distance (pc)                     & \multicolumn{2}{c}{928$\pm$36}                                \\
\enddata
\tablenotetext{a}{Potential for the inner critical surface.} 
\tablenotetext{b}{Value at 0.25 phase.} 
\tablenotetext{c}{Mean volume radius.} 
\end{deluxetable}

\begin{deluxetable}{lrccrc}
\tablewidth{0pt}
\tablecaption{Results of the multiple frequency analysis for CW Cep$\rm ^a$ }
\tablehead{
             & \colhead{Frequency}    & \colhead{Amplitude} & \colhead{Phase} & \colhead{S/N$\rm ^b$} & \colhead{Remark}               \\
             & \colhead{(day$^{-1}$)} & \colhead{(mmag)}    & \colhead{(rad)} &                &
}
\startdata                                                                                             
$f_{1}$      & 0.36642$\pm$0.00047    & 0.657$\pm$0.077     & 0.14$\pm$0.35   & 14.52          & $f_{\rm orb}$                         \\
$f_{2}$      & 2.72647$\pm$0.00051    & 0.498$\pm$0.064     & 2.03$\pm$0.37   & 13.43          & norm$\rm ^c$                          \\
$f_{3}$      & 5.34052$\pm$0.00055    & 0.330$\pm$0.045     & 4.36$\pm$0.40   & 12.47          & norm$\rm ^c$                          \\
$f_{4}$      & 0.30517$\pm$0.00164    & 0.188$\pm$0.077     & 4.25$\pm$1.20   &  4.18          & norm$\rm ^c$                          \\
$f_{5}$      & 0.73181$\pm$0.00114    & 0.265$\pm$0.076     & 2.10$\pm$0.83   &  6.01          & 2$f_{\rm orb}$                        \\
$f_{6}$      & 2.78871$\pm$0.00133    & 0.191$\pm$0.064     & 2.39$\pm$0.98   &  5.13          & norm$\rm ^c$                          \\
$f_{7}$      & 5.40878$\pm$0.00096    & 0.188$\pm$0.045     & 3.78$\pm$0.70   &  7.16          & norm$\rm ^c$                          \\
$f_{8}$      & 0.17869$\pm$0.00145    & 0.210$\pm$0.076     & 5.34$\pm$1.06   &  4.71          & norm$\rm ^c$                          \\
$f_{9}$      & 0.25498$\pm$0.00154    & 0.200$\pm$0.077     & 0.14$\pm$1.13   &  4.45          & 2$f_4$$-$$f_{\rm orb}$                \\ 
$f_{10}$     & 0.40455$\pm$0.00209    & 0.148$\pm$0.077     & 0.86$\pm$1.53   &  3.27          & 2$f_{\rm orb}$$-$$f_4$, norm$\rm ^c$  \\  
$f_{11}$     & 5.04438$\pm$0.00122    & 0.159$\pm$0.049     & 1.03$\pm$0.90   &  5.60          & $f_7$$-$$f_{\rm orb}$, norm$\rm ^c$   \\  
$f_{12}$     & 5.07951$\pm$0.00127    & 0.151$\pm$0.048     & 4.99$\pm$0.93   &  5.39          & norm$\rm ^c$                          \\
$f_{13}$     & 5.29434$\pm$0.00137    & 0.134$\pm$0.046     & 4.01$\pm$1.00   &  4.99          & norm$\rm ^c$                          \\
\enddata                                                                                                                           
\tablenotetext{a}{Frequencies are listed in order of detection. }
\tablenotetext{b}{Calculated in a range of 5 day$^{-1}$ around each frequency. }
\tablenotetext{c}{Also detected in the light residuals from the normal points. }
\end{deluxetable}

\end{document}